\documentclass[onecolumn,preprint,prd,showpacs,showkeys]{revtex4}%
\usepackage{amsfonts}
\usepackage{amsmath}
\usepackage{amssymb}
\usepackage{graphicx}%
\providecommand{\U}[1]{\protect\rule{.1in}{.1in}}

\begin{document}
\title{Acceleration effects on atomic clocks}
\author{F. Dahia}
\affiliation{Dep. of Physics, Univ. Fed. da Para\'{\i}ba, Jo\~{a}o Pessoa, Para\'{\i}ba, Brazil}
\author{P. J. Felix da Silva}
\affiliation{Dep. of Physics, Univ. Fed. de Campina Grande, Sum\'{e}, Para\'{\i}ba, Brazil.}
\altaffiliation{Dep. of Physics, Univ. Fed. da Para\'{\i}ba, Jo\~{a}o Pessoa, Para\'{\i}ba, Brazil}

\begin{abstract}
We consider a free massive particle inside a box which is dragged by Rindler
observers. Admitting that the particle obeys the Klein-Gordon equation, we
find the frequencies of the stationary states of this system. Transitions
between the stationary states are employed to set a standard frequency for a
toy atomic clock. Comparing the energy spectrum of the accelerated system with
the energy spectrum of an identical system in an inertial frame, we determine
the influence of the instantaneous acceleration on the rate of atomic clocks.
We argue that our result does not violate the clock hypothesis.

\end{abstract}
\maketitle

\subsection{Introduction}

According to the Theory of Relativity the flux of time depends on the motion
state of observers. The Lorentz transformations allow us to compare
measurements of the elapsed time between events when they are performed by
inertial observers. The predicted time dilation has been confirmed
experimentally in different contexts - Doppler shift of the energy spectrum
produced by atomic beams, lifetime of unstable particles, rate of atomic
clocks - and with an increasingly high accuracy \cite{dilation,IS}. However,
the temporal rhythm of an accelerated clock cannot be deduced from the two
fundamental postulates of the Theory of Relativity. Indeed, the discussion of
this issue demands an additional assumption. The \textit{clock hypothesis},
according to which the rate of a clock is not affected either by its
instantaneous acceleration or higher order derivatives of velocity, is usually
adopted in this context. Although it is widely accepted (for a critical
approach, see \cite{critics}), there are speculations about empirical
implications of alternative hypotheses \cite{accel,alternatives}.

The physical basis of the clock hypothesis is the so-called principle of
locality that establishes a local equivalence between an accelerated observer
and an instantaneously comoving inertial observer \cite{locality}. The idea is
that both observers momentarily share the same position and velocity and,
therefore, have the same physical state from the viewpoint of classical
mechanics. Hence, those observers would be locally equivalent, implying,
according to the principle of locality, that they would have the same notion
of time, instantaneously. Mathematically this is translated into the
well-known relation between the proper time $(d\tau)$ of the accelerated
observer and the time $(dt)$ as measured in an inertial frame: $d\tau
=dt/\gamma$, where $\gamma$ is the Lorentz factor.

So far the clock hypothesis has been confirmed experimentally. Measurements of
the lifetime of unstable particles moving in circular orbits or when they are
submitted to longitudinal acceleration are in agreement with the clock
hypothesis \cite{centripetal,longitudinal}. Within the accuracy of the
instruments, these tests found no evidence of the influence of the
acceleration on the decay rate, despite those particles have been subjected to
very high acceleration: 10$^{18}g$ (for the centripetal acceleration
\cite{centripetal}) and $10^{15}g$ (for the average longitudinal acceleration,
with peaks of $a=10^{22}g$ \cite{longitudinal}), where $g$ is the Earth
gravitational acceleration.

On the other hand, it is worthy of mention that the principle of locality
relies on the definition of physical state established by classical mechanics.
However, as Nature has a quantum character, then, the locality principle has a
conceptual limitation \cite{accel,limitation,newformula}. Because of this, it
has been suggested that the rate of any actual clock would be influenced by
its instantaneous acceleration or could be even affected by its past world
line \cite{accel,limitation}. Thus the relation between $d\tau$ and $dt$
should be modified when the wave behavior of system is taken into account. For
instance, based on the study of the lifetime of artificial muonic particles in
a circular motion \cite{eisele}, it was proposed that, regarding the muon
proper time, the new relation would be: $d\tau=\left(  1+2/3\left(
\lambdabar/%
\mathcal{L}%
\right)  ^{2}\right)  dt/\gamma$, where $\lambda$ is the muon Compton
wavelength, $\lambdabar=\lambda/2\pi$ and $%
\mathcal{L}%
=c^{2}/a$ defines a characteristic length scale where the particle state
changes significantly \cite{newformula}. Under the conditions in which the
muon experiment was performed $a\sim10^{18}g$ \cite{centripetal}, which
corresponds to $%
\mathcal{L}%
\simeq1%
\operatorname{cm}%
$. The muon Compton wavelength is of the order of $10^{-14}%
\operatorname{m}%
$, then $\left(  \lambdabar/%
\mathcal{L}%
\right)  ^{2}$ is less than $10^{-25}$. This small correction could not be
detected within the accuracy of that experiment. So these effects of the
acceleration cannot be ruled out and the question remains open.

In this paper, we want to address this issue investigating the behavior of an
accelerated atomic clock. Actually, for the sake of simplicity, our atomic
clock is described by a toy model which consists of a particle in a box which
obeys the Klein-Gordon equation. The acceleration of the box is implemented,
within our scheme, assuming that the walls are dragged by Rindler observers.
Solving the Klein-Gordon equation in the Rindler frame and imposing the
appropriate boundary conditions, we determine the frequency of the stationary
states of the system. Comparing this spectrum with the spectrum of an
identical system in an inertial system, we can determine a relation between
the ticking rate of the accelerated clock and the rate of an inertial atomic clock.

\section{Accelerated atomic clock}

The world line of an uniformly accelerated observer in a Minkowski spacetime
is described, in terms of the rectangular coordinates $\left(  t,x\right)  $
of an inertial frame, by the following parametric curve:%
\begin{align}
t\left(  \tau\right)   &  =\frac{c}{a}\sinh\left(  a\tau/c\right)
,\label{time}\\
x\left(  \tau\right)   &  =\frac{c^{2}}{a}\cosh\left(  a\tau/c\right)  ,
\label{x}%
\end{align}
where $\tau$ is the observer's proper time (defined in terms of the length of
the curve) and $a$ is its proper acceleration. The world line corresponds to a
hyperbola in the spacetime diagram of the inertial frame. For the sake of
simplicity, we are considering a spacetime of $\left(  1+1\right)  $-dimensions.

The accelerated observer can construct a coordinate system adapted to its
motion by using the locality principle. At a particular instant of time $\tau
$, the observer, with the help of the instantaneously comoving inertial frame
$S_{\tau}$, determines the events which are simultaneous. All of them are
labelled with the same temporal coordinate $\tau$. The spatial coordinate,
$\xi$, of these simultaneous events are established by using the spatial
coordinate of $S_{\tau}$. From this definition, it follows that the
transformation between $\left(  t,x\right)  $ and the Rindler coordinates
$\left(  \tau,\xi\right)  $ is given by:%
\begin{align}
ct  &  =\left(  c^{2}/a+\xi\right)  \sinh\left(  a\tau/c\right)  ,\\
x  &  =\left(  c^{2}/a+\xi\right)  \cosh\left(  a\tau/c\right)  .
\end{align}
It is well known that this transformation is defined only in a region
corresponding to the right hand side of the light cone whose vertex is at the
origin of the inertial frame. Each coordinate line $\xi=constant$ is an
hyperbola in the spacetime diagram of $S$ (in $\left(  t,x\right)
$-coordinates) and represents the world line of an uniformly accelerated
observer with proper acceleration equals to $a/\left(  1+a\xi/c^{2}\right)  $
\cite{patricio}. The set of these observers constitutes the Rindler frame. The
Minkowski metric written in the Rindler coordinates assumes the following
form:%
\begin{equation}
ds^{2}=-c^{2}\left(  1+a\xi/c^{2}\right)  ^{2}d\tau^{2}+d\xi^{2}.
\end{equation}

Consider now a particle with rest mass $m$ confined in a box which is dragged
by Rindler observers. This means the walls are found at rest with respect to
the Rindler frame. Let us assume that the walls are localized at positions
$\xi_{1}$ and $\xi_{1}+\ell$ (where $\ell$ is the proper length of the box as
measured in the Rindler frame), around the central Rindler observer (i.e., the
observer at $\xi=0$). Inside the box, the particle is free. Admitting that its
behavior is governed by the Klein-Gordon equation, then, in the Rindler
coordinates, the particle's wave function $\phi\left(  \tau,\xi\right)  $
satisfies the following equation:%
\begin{equation}
-\frac{1}{c^{2}\left(  1+a\xi/c^{2}\right)  ^{2}}\frac{\partial^{2}\phi
}{\partial\tau^{2}}+\frac{1}{\left(  1+a\xi/c^{2}\right)  }\frac{\partial
}{\partial\xi}\left(  \left(  1+a\xi/c^{2}\right)  \frac{\partial\phi
}{\partial\xi}\right)  -\left(  \frac{mc}{\hbar}\right)  ^{2}\phi=0.
\end{equation}
At this point it is important to emphasize that although the interpretation of
$\phi$ as a wave function has some limitations, this does not affect the
purpose of our discussion, since we are dealing with a toy model and we are
basically concerned with the mathematical problem of finding the stationary
states of KG equation and their respective frequencies.

The solution can be written as $\phi=e^{-i\omega\tau}\psi$, where $\psi$ is
independent of $\tau$ and satisfies the modified Bessel equation:%
\begin{equation}
\rho^{2}\frac{d^{2}\psi}{d\rho^{2}}+\rho\frac{d\psi}{d\rho}-\left[  \rho
^{2}-\frac{\omega^{2}c^{2}}{a^{2}}\right]  \psi=0, \label{Bessel}%
\end{equation}
where $\rho=(mc/\hbar)\left(  c^{2}/a+\xi\right)  $. Here we can identify
three length scales: the particle's reduced Compton wavelength $\lambdabar
=\hbar/mc$, the size $\ell$ of the box and the acceleration length $%
\mathcal{L}%
=c^{2}/a$. The general solution of equation (\ref{Bessel}) is a linear
combination of the modified Bessel function of the first kind $I_{\nu}\left(
\rho\right)  $ and of second kind $K_{\nu}\left(  \rho\right)  $ with the
order $\nu=i\left(  \omega c/a\right)  $. As the order is pure imaginary,
$I_{\nu}\left(  \rho\right)  $ is complex on the positive real axis and,
hence, a better companion for $K_{\nu}\left(  \rho\right)  $ is the function
$L_{\nu}\left(  x\right)  =$ $\left[  I_{\nu}\left(  x\right)  +I_{-\nu
}\left(  x\right)  \right]  /2$ which is also a real function for $x>0$ in the
case of a pure imaginary order \cite{fabijonas,dunster}. Thus, the general
solution can be appropriately written as $\psi\left(  \rho\right)
=C_{1}L_{\nu}\left(  \rho\right)  +C_{2}K_{\nu}\left(  \rho\right)  $, where
$C_{1}$ and $C_{2}$ are arbitrary complex numbers. The stationary states in
the box and their corresponding frequencies are determined by the boundary
conditions: $\psi=0$ at the points $\rho_{1}=(mc/\hbar)\left(  c^{2}/a+\xi
_{1}\right)  $ and $\rho_{2}=(mc/\hbar)\left(  c^{2}/a+\xi_{1}+\ell\right)  $.
At least one of the coefficients must be nonzero, otherwise the wave function
would be identically null. It happens that there exists a non-trivial solution
for $C_{1}$ and $C_{2}$ only if the elements $L_{\nu}\left(  \rho_{1}\right)
$, $K_{\nu}\left(  \rho_{1}\right)  $, $L_{\nu}\left(  \rho_{2}\right)  $ and
$K_{\nu}\left(  \rho_{2}\right)  $ constitute a matrix with null determinant.
This condition is equivalent to the equation:%
\begin{equation}
K_{\frac{i\omega c}{a}}\left(
\mathcal{L}%
/\lambdabar+\xi_{1}/\lambdabar\right)  L_{\frac{i\omega c}{a}}\left(
\mathcal{L}%
/\lambdabar+\left(  \xi_{1}+\ell\right)  /\lambdabar\right)  =K_{\frac{i\omega
c}{a}}\left(
\mathcal{L}%
/\lambdabar+\left(  \xi_{1}+\ell\right)  /\lambdabar\right)  L_{\frac{i\omega
c}{a}}\left(
\mathcal{L}%
/\lambdabar+\xi_{1}/\lambdabar\right)  . \label{bc}%
\end{equation}

We want to analyze the above equation in the small acceleration regime. For
this purpose, it is appropriate to consider the asymptotic expansion of the
modified Bessel functions \cite{fabijonas,dunster}. Taking the first terms, we
have \cite{fabijonas}:%

\begin{align}
e^{\alpha\pi/2}K_{i\alpha}\left(  \alpha x\right)   &  \simeq\frac{1}{\sqrt
{2}}\left[  \cos\left(  \alpha\theta\left(  x\right)  +\frac{\pi}{4}\right)
\Pi_{1}+\sin\left(  \alpha\theta\left(  x\right)  +\frac{\pi}{4}\right)
\Pi_{2}\right]  ,\label{K}\\
e^{-\alpha\pi/2}L_{i\alpha}\left(  \alpha x\right)   &  \simeq-\frac{1}%
{2\sqrt{2}\pi}\left[  -\cos\left(  \alpha\theta\left(  x\right)  +\frac{\pi
}{4}\right)  \Pi_{2}+\sin\left(  \alpha\theta\left(  x\right)  +\frac{\pi}%
{4}\right)  \Pi_{1}\right]  , \label{L}%
\end{align}
where $\alpha=\omega c/a$, $\theta\left(  x\right)  =\sqrt{1-x^{2}}-\ln\left(
\left(  1+\sqrt{1-x^{2}}\right)  /x\right)  $ and approximate expressions for
$\Pi_{1}$ and $\Pi_{2}$ are:%

\begin{equation}
\Pi_{1}\sim2\sqrt{\frac{\pi}{\alpha}}\left[  1-x^{2}\right]  ^{-1/4}%
\overset{\infty}{\underset{s=0}{\sum}}\frac{V_{2s}(i\left[  1-x^{2}\right]
^{-1/2})}{\alpha^{2s}},
\end{equation}

\begin{equation}
\Pi_{2}\sim-2\sqrt{\frac{\pi}{\alpha}}\left[  1-x^{2}\right]  ^{-1/4}%
\overset{\infty}{\underset{s=0}{\sum}}\frac{iV_{2s+1}(i\left[  1-x^{2}\right]
^{-1/2})}{\alpha^{2s+1}},
\end{equation}
where%
\begin{equation}
V_{0}\left(  q\right)  =1,V_{s+1}\left(  q\right)  =\frac{1}{2}q^{2}\left(
q^{2}+1\right)  V_{s}^{\prime}\left(  q\right)  +\frac{1}{8}\int_{0}^{q}%
V_{s}\left(  t\right)  \left(  1+5t^{2}\right)  dt,
\end{equation}
for $s=0,1,2,...$\cite{dunster,fabijonas}$.$ By using this approximation in
the equation (\ref{bc}), we obtain the frequency, or speaking loosely, we find
the energy $\left(  \hbar\omega\right)  $ of the stationary states. In the
second order of $a$, the spectrum of positive energy as measured in the
Rindler frame is given by:
\begin{align}
E_{n}=E_{n}^{\left(  0\right)  }  &  \left\{  1+\left(  \frac{\ell}{2%
\mathcal{L}%
}+\frac{\xi_{1}}{%
\mathcal{L}%
}\right)  +\left[  -\frac{1}{12}+\frac{1}{8n^{2}\pi^{2}}\left(  \frac
{1}{1+n^{2}\pi^{2}\lambdabar^{2}/\ell^{2}}\right)  \right]  \left(  \frac
{\ell}{%
\mathcal{L}%
}\right)  ^{2}\right. \nonumber\\
&  \left.  +\frac{1}{8}\left(  -\frac{5}{n^{4}\pi^{4}}+\frac{1}{3n^{2}\pi^{2}%
}\right)  \left(  \frac{\ell^{2}}{%
\mathcal{L}%
\lambdabar}\right)  ^{2}\right\}  , \label{energy}%
\end{align}
where $E_{n}^{\left(  0\right)  }=\left[  m^{2}c^{4}+\left(  n^{2}\pi^{2}%
\hbar^{2}c^{2}\right)  /\ell^{2}\right]  ^{1/2}$ is the energy of the quantum
level $n$ of an identical system (a mass $m$ in a box of proper length $\ell$)
at rest in an inertial frame. The equation (\ref{energy}) is valid for
$n^{2}>\left(  1/\pi^{2}\right)  \ell^{3}/\left(
\mathcal{L}%
\lambdabar^{2}\right)  $ (see appendix A). Note that the linear term relative
to the acceleration $a$ depends on the position of the box with respect the
central Rindler observer. In our analogy with an actual atom, let us assume
that the path of this observer (at $\xi=0$) corresponds to the trajectory of
the atomic nucleus. Thus, a symmetric configuration of the walls around the
nucleus (which seems to be the most natural choice, otherwise the atom would
have a spontaneous electric dipole) corresponds to $\xi_{1}=-\ell/2$. In this
case, the linear term vanishes and, therefore, the leading correction is
quadratic. If we write the length $\ell$ of the box, the wavelength
$\lambdabar$ and the acceleration $a$ in terms of Bohr radius $\left(
r_{0}\right)  $, the electron mass $\left(  m_{e}\right)  $ and $g$,
respectively, we find the following estimates:%
\begin{align}
\frac{\ell}{%
\mathcal{L}%
}  &  \sim10^{-27}\left(  \frac{\ell}{r_{0}}\right)  \left(  \frac{a}%
{g}\right)  ,\\
\frac{\lambdabar}{%
\mathcal{L}%
}  &  \sim10^{-29}\left(  \frac{m_{e}}{m}\right)  \left(  \frac{a}{g}\right)
.
\end{align}
For $a=10^{18}g$, the quadratic correction produces a relative shift of the
order of $10^{-18}\left(  \ell/r_{o}\right)  ^{2}$.

In analogy with actual atomic processes, let us assume that, in a transition
between two stationary states of our system, a quantum of some field with a
well-defined frequency is emitted. Some device endowed with a counter of the
cycles of the standard frequency defines our toy clock.

Now let us consider a transition from a certain state $f$ to the quantum level
$i.$ The emitted quantum has a frequency $\omega_{fi}$ that is employed as the
standard frequency of the accelerated clock. When measured in terms of the
parameter $\tau$, the frequency $\omega_{fi}=\left(  E_{f}-E_{i}\right)
/\hbar$ may be determined from equation (\ref{energy}). On its turn, for an
identical system at rest in the inertial system $S$, the frequency of the
emitted quantum in the corresponding transition from state $f$ to $i$ is given
by $\omega_{fi}^{0}=$ $\left(  E_{f}^{0}-E_{i}^{0}\right)  /\hbar$. Note that
$\omega_{fi}^{0}$ is measured in terms of the time coordinate of the frame
$S$. Having this in mind, now consider two close events $E_{1}$ and $E_{2}$ on
the path of the central accelerated observer, with labels $\tau_{1}$ and
$\tau_{2}$ respectively. If $\Delta T$ is the number of oscillations of the
standard wave with frequency $\omega_{fi}$ that happen during the elapsed time
between $E_{1}$ and $E_{2}$, then:%
\begin{equation}
\Delta T=\left(  \tau_{2}-\tau_{1}\right)  \omega_{fi}/2\pi. \label{T}%
\end{equation}
On the other hand, from the point of view of $S$, during the interval
$\Delta\tau=\tau_{2}-\tau_{1}$, the number of cycles of the inertial atomic
clock running in the standard frequency $\omega_{fi}^{0}$ is:
\begin{equation}
\Delta t=\left(  c/a\right)  [\sinh\left(  a\tau_{2}/c\right)  -\sinh\left(
a\tau_{1}/c\right)  ]\omega_{fi}^{0}/2\pi, \label{t}%
\end{equation}
according to equation (\ref{time}). Therefore, in the limit $\Delta
\tau\rightarrow0$, the instantaneous ratio between the ticking rates of the
clocks is:%
\begin{equation}
\frac{dT}{dt}=\sqrt{1-v^{2}/c^{2}}\left(  \frac{\Delta E_{fi}}{\Delta
E_{fi}^{0}}\right)  . \label{ratioTt}%
\end{equation}
where $v$ is the relative velocity between the observers at the instant
$\tau_{1}$(see appendix\ B). The correction factor depends on the frequency of
the atomic clock, i.e., on the particular transition $\left(  f\rightarrow
i\right)  $ that is chosen to set the standard frequency. However, in
transitions with higher quantum numbers, the relation reduces to:
\begin{equation}
\frac{dT}{dt}=\sqrt{1-v^{2}/c^{2}}\left[  1-\frac{1}{12}\left(  \frac{\ell}{%
\mathcal{L}%
}\right)  ^{2}\right]  . \label{ticking}%
\end{equation}
Therefore the rate of an accelerated atomic clock depends on its instantaneous
acceleration. Moreover the leading term of corrections is quadratic with
respect to the acceleration. This result is similar to that suggested in Refs.
\cite{eisele,newformula}, which predicts a correction of the order of $\left(
\lambdabar/%
\mathcal{L}%
\right)  ^{2}$ for the muon lifetime. Nevertheless, it is important to stress
that, in comparison with the study of the muon proper time \cite{eisele}, our
result differs in some aspects. In Ref. \cite{eisele}, the influence of a
magnetic field on the decay rate of muons was determined. As this magnetic
field is responsible to keep the muons in circular motion, its influence can
be written in terms of the \textit{centripetal} acceleration \cite{newformula}%
. On the other hand, here we have studied the influence of a
\textit{longitudinal} acceleration over the energy levels of an atom.
Therefore, the systems are physically distinct and are accelerated in
different ways. Besides, the approaches are also different: Ref. \cite{eisele}
follows the second quantization approach, which is the appropriate formalism
to deal with corrections of Compton wavelength order, since it is expected
that, in this scale, the quantum field theory effects of particle creation and
annihilation take place; In contrast, our method is based on the first
quantization approach. We find that the dominant correction does not depend on
the Compton wavelength of the particle, but on the size $\ell$ of the atom,
justifying, in this way, the use of the much simpler formalism of the first
quantization. So, the result of Ref. \cite{eisele} and ours are independent
and, let us say, complementary since they contemplate different aspects of the problem.

The acceleration effect on the rhythm of atomic clocks may be much greater
than the effect on the muon lifetime, since the correction depends on $\left(
\ell/%
\mathcal{L}%
\right)  ^{2}$ rather than $\left(  \lambda/%
\mathcal{L}%
\right)  ^{2}$. Indeed, if the size of the box is of the order of the Bohr
radius $\left(  \gtrsim10^{-11}%
\operatorname{m}%
\right)  $, then $\left(  \ell/%
\mathcal{L}%
\right)  ^{2}$ is slightly greater than $10^{-18}$ when $a=10^{18}g$. Thus the
acceleration effects on the proper time of an atomic clock will be, in a
conservative estimate, 10$^{6}$ times greater than those that were predicted
for a muon circulating with the same acceleration \cite{eisele,decay law}.
Furthermore, the fact that the systems are accelerated in different ways may
have experimental implications. In the study of longitudinally accelerated
particle, the acceleration reaches peaks of $10^{22}g$ \cite{longitudinal}. In
this order of magnitude, the effect of acceleration will be approximately
$\left(  \ell/%
\mathcal{L}%
\right)  ^{2}\sim10^{-10}$, which is close to the current accuracy
($\sim10^{-9}$) of empirical tests of the time dilation \cite{IS}.

The instantaneous acceleration influences the ticking rate of the clock and,
as a consequence, it also affects the Doppler shift. According to
(\ref{energy}), in transitions with higher quantum numbers, the frequency of
the emitted quantum by the accelerated source is $\omega_{E}^{0}\left(
1-\ell^{2}/12%
\mathcal{L}%
^{2}\right)  $, where $\omega_{E}^{0}$ is the frequency for the same
transition when it happens in an inertial system. If the accelerated source
emits forward and backward signals, then the frequency as measured by an
inertial receiver is
\begin{equation}
\omega_{R}^{0}=\gamma\left(  1\pm\frac{v}{c}\right)  \left[  1\pm\frac
{\lambda_{0}}{2%
\mathcal{L}%
}+\frac{1}{12}\left(  \frac{\lambda_{0}}{%
\mathcal{L}%
}\right)  ^{2}-\frac{1}{12}\left(  \frac{\ell}{%
\mathcal{L}%
}\right)  ^{2}\right]  \omega_{E}^{0},
\end{equation}
where $v$ is the relative velocity at the moment of emission and $\lambda_{0}$
is the wavelength of the corresponding signal emitted in an inertial frame
(see appendix C). The signs $+$ or $-$ are valid when the observers are
approaching or receding respectively. The acceleration modifies the
Doppler-shift formula in two different ways: the term $\lambda_{0}/%
\mathcal{L}%
$ is associated with the variation of the source's velocity during a complete
cycle; while $\left(  \ell/%
\mathcal{L}%
\right)  ^{2}$ is a new contribution that arises due to the change of the
ticking rate of clock caused by its acceleration.

\section{ Discussion and Conclusion}

We showed that the rate of an accelerated clock is affected by its
instantaneous acceleration, however this result should not be seen as a
violation of the clock hypothesis. Actually the Rindler frame is built with
the help of this hypothesis and even the usual form the KG equation assumes in
the accelerated frame is somehow based on the principle of locality. The
influence of the instantaneous acceleration on the rhythm of an atomic clock
is just the expression of the fact that the internal dynamics of a system of
finite size is affected by acceleration. This is true even in the context of
the classical mechanics, as we can check by calculating the period of an
accelerated pendulum or the period of oscillations of a light beam between
accelerated mirrors \cite{classical}. In view of this, the better
interpretation of our result is that the rate of accelerated actual clock
deviates from the rate of an ideal point-like clock \cite{pointlike}. In this
sense, we can say that the equation (\ref{ticking}), actually, is compatible
with the clock hypothesis, since, according to it, the rate of atomic clock
does not depend on its instantaneous acceleration in the limit $\ell
\rightarrow0$ and $\lambdabar\rightarrow0$. Nevertheless, as $\ell$ and
$\lambdabar$ are non-null for actual physical system, the instantaneous
acceleration produces some effects on the ticking rate of accelerated clocks.
Perhaps the most important result of our study is the fact that, with the help
of this simple toy model, we can identify what are the relevant parameters
that contribute to the these effects and estimate the magnitude order of them
by using equation (\ref{ticking}). As we have seen, the influence of the
acceleration on the proper time are very tiny, however, as the experiments are
getting more accurate, it is important to take them into account in order to
make a correct interpretation of empirical data. Recently an Ives-Stiwell type
experiment has tested the time dilation factor within accuracy of 10$^{-9}$
\cite{IS}. If the same precision could be achieved in experiments involving
acceleration of the order greater than $10^{23}g$ then the acceleration
effects on the ticking rate of an atomic clock would become detectable. As we
have already mentioned, in the experiment involving longitudinal acceleration
\cite{longitudinal}, the acceleration has peaks of $10^{22}g$.

Our toy model is a very simplified system. To deal with a real atom, first we
have to devise an acceleration mechanism. In the case of a neutral atom, we
should employ a non-uniform electric field to produce a longitudinal
acceleration. Thus, in a realistic model, we have to consider the interaction
of the external field with the atomic particles whose behavior is governed by
the Dirac equation in (3+1)-dimensions. This problem is, of course, much more
complicated. The effects of the acceleration over the energy levels would
appear indirectly as consequence of the interaction between the external field
and the atom. Nevertheless, if the external field is not so strong, in
comparison with the strength of the internal interaction, the atom will remain
as a bound system, with an accelerated center of mass. The fundamental
characteristic of our model is that it represents an accelerated bound system.
Then, in this sense, a particle in an accelerated box can be consider as a toy
model for an accelerated real atom. However, we must bear in mind that this
study is a preliminary work and that our results need to be improved in light
of more realistic models.

\section{Acknowledgement}

P.J.F.S. thanks CAPES for financial support.

\section{Appendix A}

The expansion of the functions $L_{i\alpha}\left(  \alpha x\right)  $ and
$K_{i\alpha}\left(  \alpha x\right)  $ given by eqs. (\ref{K}) and (\ref{L})
are valid when the argument $\alpha x$ is lesser than the order $\alpha$, i.e,
for $x<1$ \cite{fabijonas,dunster}. To use these formulas in the equation
(\ref{bc}) we have to make $\alpha x=$ $%
\mathcal{L}%
/\lambdabar+\xi_{1}/\lambdabar$ or $\alpha x=$ $%
\mathcal{L}%
/\lambdabar+\left(  \xi_{1}+\ell\right)  /\lambdabar$, where $\xi_{1}$ is the
position of the first wall. Taking $\xi_{1}=-\ell/2$ (corresponding to a
symmetric configuration of the box) and considering that $\alpha=\omega c/a$,
we find that the maximum value of $x$ is $\left(  c/\lambdabar+\ell
a/2\lambdabar c\right)  /\omega$. Thus, the condition $x<1$ implies that
\begin{equation}
\hbar\omega>mc^{2}+\frac{1}{2}ma\ell,
\end{equation}
recalling that $\lambdabar=\hbar/mc$. As $\omega$ is the frequency of a
certain stationary state, then, $\hbar\omega$ is the corresponding energy. The
energy of the unperturbed system is given by $\left[  m^{2}c^{4}+\left(
n^{2}\pi^{2}\hbar^{2}c^{2}\right)  /\ell^{2}\right]  ^{1/2}$. Thus, in first
of order of $a$, we have the condition
\begin{equation}
n^{2}>\frac{1}{\pi^{2}}\frac{\ell^{3}}{\lambdabar^{2}%
\mathcal{L}%
} \label{n}%
\end{equation}
This condition may be interpreted in two different ways. If the system (a
particle of mass $m$ confined in the box of size $\ell$) is moving with a
known proper acceleration $a$, then the above inequality establishes what are
the quantum levels that can be used to define a standard frequency for the
atomic clock. On the other hand, if the quantum levels are chosen previously,
equation (\ref{n}) gives the maximum acceleration which is consistent with our
approximation scheme. To make some estimates, let us consider that $\ell
\simeq0.5\times10^{-10}%
\operatorname{m}%
$ (Bohr radius) and that $m$ is the electron mass, which corresponds to
$\lambda\simeq2.4\times10^{-12}%
\operatorname{m}%
$. Thus, $%
\mathcal{L}%
>10^{-7}/n^{2}$. Therefore, even for the level $n=1$, the bound on the
acceleration, $a<10^{22}g$, is not stringent.

\section{Appendix B}

According to equation (\ref{T}), $\Delta T$ is the number of tick-tacks
between the events $E_{1}$ and $E_{2}$ as measured by the accelerated clock.
On the other hand, $\Delta t$ in equation (\ref{t}) gives the number of
tick-tacks as measured by an inertial frame. The ratio between these
quantities is%
\begin{equation}
\frac{\Delta t}{\Delta T}=\left(  c/a\right)  \frac{[\sinh\left(  a\tau
_{2}/c\right)  -\sinh\left(  a\tau_{1}/c\right)  ]}{\left(  \tau_{2}-\tau
_{1}\right)  }\omega_{fi}^{0}/\omega_{fi},
\end{equation}
where we can write $\tau_{2}=\tau_{1}+\Delta\tau$. In the limit when
$\Delta\tau\rightarrow0$, we have%
\begin{equation}
\frac{dt}{dT}=\cosh\left(  a\tau_{1}/c\right)  \omega_{fi}^{0}/\omega_{fi}%
\end{equation}
From the motion equations of the accelerated observer, eqs. (\ref{time}) and
(\ref{x}), we can verify that the instantaneous velocity of the accelerated
observer with respect the inertial system is $v=dx/dt=c\tanh\left(  a\tau
_{1}/c\right)  $. Thus, writing $\cosh\left(  a\tau_{1}/c\right)  $ as
$1/\sqrt{1-(v/c)^{2}}$, we obtain equation (\ref{ratioTt}).

\section{Appendix C}

Consider that the accelerated observer emits signals at the time $\tau_{1}$
and $\tau_{2}$. The coordinates of the emission events can be determined from
equations (\ref{time}) and (\ref{x}). The signals travel at the light velocity
and reach an inertial observer fixed at $x=0$ at instant $t_{1}$ and $t_{2}$
respectively. Therefore, we can write%
\begin{align}
t_{1}-t\left(  \tau_{1}\right)   &  =\frac{1}{c}\left(  x\left(  \tau
_{1}\right)  -0\right)  ,\\
t_{2}-t\left(  \tau_{2}\right)   &  =\frac{1}{c}\left(  x\left(  \tau
_{2}\right)  -0\right)  .
\end{align}
It follows that%
\begin{equation}
t_{2}-t_{1}=t\left(  \tau_{2}\right)  -t\left(  \tau_{1}\right)  +\frac{1}%
{c}\left(  x\left(  \tau_{2}\right)  -x\left(  \tau_{1}\right)  \right)
\end{equation}
If we write $\tau_{2}=\tau_{1}+\Delta\tau$ and by using the equations
(\ref{time}) and (\ref{x}), we find
\begin{equation}
\Delta t=\frac{c}{a}\exp\left(  a\tau_{1}/c\right)  \cdot\left[  \exp\left(
a\Delta\tau/c\right)  -1\right]  ,
\end{equation}
where $\Delta t=t_{2}-t_{1}$. Now, expanding the above equation up to the
second order of $a$, we obtain%
\begin{equation}
\Delta t=\exp\left(  a\tau_{1}/c\right)  \cdot\left[  1+\frac{1}{2}\left(
c\Delta\tau/%
\mathcal{L}%
\right)  +\frac{1}{6}\left(  c\Delta\tau/%
\mathcal{L}%
\right)  ^{2}\right]  \Delta\tau
\end{equation}
If $\Delta\tau$ is the wave period as measured by the accelerated observer,
then $\Delta t$ is the period of the wave received by the inertial observer.
Therefore, the relation between the frequencies is:%
\begin{equation}
\omega_{R}^{0}=\exp\left(  -a\tau_{1}/c\right)  \cdot\left[  1-\frac{1}%
{2}\left(  c\Delta\tau/%
\mathcal{L}%
\right)  +\frac{1}{12}\left(  c\Delta\tau/%
\mathcal{L}%
\right)  ^{2}\right]  \omega_{E}^{\prime}.
\end{equation}
As we have already mentioned, the frequency emitted is $\omega_{E}^{\prime}=$
$\omega_{E}^{0}\left(  1-\ell^{2}/12%
\mathcal{L}%
^{2}\right)  $. Therefore, recalling that $\Delta\tau=2\pi/\omega_{E}^{\prime
}$, we can write $\left(  c\Delta\tau/%
\mathcal{L}%
\right)  $, up to the second order of $a$, as $\left(  \lambda_{0}/%
\mathcal{L}%
\right)  $, where $\lambda_{0}=2\pi c/\omega_{E}^{0}$ would be the wavelength
of the signal if it had been emitted by an inertial frame. On its turn, from
the fact that instantaneous relative velocity is $v=c\tanh\left(  a\tau
_{1}/c\right)  $, we can write $\exp\left(  -a\tau_{1}/c\right)  $ as
$\gamma\left(  1-\left(  v/c\right)  \right)  $. If follows that
\begin{equation}
\omega_{R}^{0}=\gamma\left(  1-\frac{v}{c}\right)  \left[  1-\frac{1}%
{2}\left(  \lambda_{0}/%
\mathcal{L}%
\right)  +\frac{1}{12}\left(  \lambda_{0}/%
\mathcal{L}%
\right)  ^{2}-\frac{1}{12}\left(  \ell/%
\mathcal{L}%
\right)  ^{2}\right]  \omega_{E}^{0}%
\end{equation}
For $\tau_{1}>0$, the velocity is positive and the accelerated observer is
running away from the inertial receiver located at $x=0$.

Now, if we consider a receiver at a certain fixed position $x=D$ at the right
hand side of the accelerated observer, we can show, following the same
reasoning, that%
\begin{equation}
\omega_{R}^{0}=\gamma\left(  1+\frac{v}{c}\right)  \left[  1+\frac{1}%
{2}\left(  \lambda_{0}/%
\mathcal{L}%
\right)  +\frac{1}{12}\left(  \lambda_{0}/%
\mathcal{L}%
\right)  ^{2}-\frac{1}{12}\left(  \ell/%
\mathcal{L}%
\right)  ^{2}\right]  \omega_{E}^{0}%
\end{equation}
In this case, for $\tau_{1}>0$, the observers are approaching.

\end{document}